\documentclass[apj]{emulateapj}

\usepackage{color}

\begin{document}

\title{Attempt to a Unified model for the gamma-ray emission of supernova 
remnants}

\author{Qiang Yuan$^1$, Siming Liu$^{2}$ and Xiao-Jun Bi$^1$}

\affil{$^1$Key Laboratory of Particle Astrophysics, Institute of High
Energy Physics, Chinese Academy of Sciences, Beijing 100049, China\\
$^2${Key Laboratory of Dark Matter and Space Astronomy, Purple
Mountain Observatory, Chinese Academy of Sciences, Nanjing 210008, China}
}

\begin{abstract}

Shocks of supernova remnants (SNRs) are important (and perhaps the 
dominant) agents for production of the Galactic cosmic rays. Recent 
$\gamma$-ray observations of several SNRs have made this case more 
compelling. However, these broadband high-energy measurements also 
reveal a variety of spectral shape demanding more comprehensive 
modeling of emissions from SNRs. According to the locally observed 
fluxes of cosmic ray protons and electrons, the electron-to-proton
number ratio is known to be about $1\%$. Assuming such a ratio is 
universal for all SNRs and identical spectral shape for all kinds 
of accelerated particles, we propose a unified model that ascribes 
the distinct $\gamma$-ray spectra of different SNRs to variations 
of the medium density and the spectral difference between cosmic 
ray electrons and protons observed at Earth to transport effects. 
For low density environments, the $\gamma$-ray emission is 
inverse-Compton dominated. For high density environments like systems 
of high-energy particles interacting with molecular clouds, the
$\gamma$-ray emission is $\pi^0$-decay dominated. The model predicts 
a hadronic origin of $\gamma$-ray emission from very old remnants 
interacting mostly with molecular clouds and a leptonic origin for 
intermediate age remnants whose shocks propagate in a low density 
environment created by their progenitors via e.g., strong stellar 
winds. These results can be regarded as evidence in support of
the SNR-origin of the Galactic cosmic rays.

\end{abstract}


\section{Introduction}

Nearly a century after the discovery of cosmic rays (CRs), their
origin remains one of the biggest fundamental questions in the
fields of high energy physics and astrophysics. It is widely
believed that the supernova remnants (SNRs) are one of the most
probable candidate sources of the Galactic CRs below the so-called
``knee'' \citep{1985Natur.314..515B,
2001JPhG...27..941E,2001JPhG...27..959E,2001JPhG...27.1709E,
2002JPhG...28.2329E,2005JPhG...31R..95H,2008JCAP...01..018K}.
However, there are currently no observations that can directly
verify such a conjecture.

The multi-wavelength observations of SNRs, especially the high energy
$\gamma$-rays from the ground-based atmospheric Cerenkov telescope
arrays and space-based telescopes, provide powerful tools to probe the
particle acceleration mechanism in SNRs \citep{2002Natur.416..823E,
2004Natur.432...75A,2007Natur.449..576U,2007ApJ...657L..25T,
2008ApJ...679L..85T,2010ApJ...712..790T,2012PhRvD..85h3008N}.
However, whether the nature of these $\gamma$-ray emission from SNRs
is predominantly hadronic or leptonic is still a matter of debate
\citep[e.g.,][]{2006A&A...449..223A,2006A&A...451..981B,2008MNRAS.386L..20B,
2008ApJ...683L.163L,2008NewA...13...73P,2009MNRAS.392..240M,
2009MNRAS.392..925F,2010A&A...517L...4F,2010ApJ...712..287E,
2010ApJ...708..965Z,2010A&A...511A..34B,2011ApJ...735..120Y}.
That is to say, SNRs are known particle accelerators, but it is
not clear whether they dominate the observed Galactic CR flux,
especially for nuclei, on the Earth.

The observational $\gamma$-ray spectra of SNRs also show significant
diversity. Recent Fermi observations of the young SNR RX J1713.7-3946
shows a very hard spectrum in GeV energy range, $1.50\pm0.11$, which
implies an inverse-Compton (IC) origin of the $\gamma$-rays
\citep[][see \cite{2012ApJ...744...71I} for an alternative explanation]
{2011ApJ...734...28A}. On the other hand, for all of the SNRs
interacting with molecular clouds (MCs), the GeV-TeV $\gamma$-ray spectra
are generally very soft and seem to better agree with the $\pi^0$-decay
model, although the model of bremsstrahlung emission from electrons can
not be ruled out \citep{2010Sci...327.1103A,2009ApJ...706L...1A,
2010ApJ...712..459A,2010ApJ...710L..92A,2010ApJ...718..348A,
2010ApJ...722.1303A,2012ApJ...744...80A,2010A&A...516L..11G}.
It is natural to ask whether there is a common understanding of these
$\gamma$-ray signatures of the SNRs.

In this paper we illustrate that the locally observed CRs and the
$\gamma$-ray emission of SNRs can be naturally understood in a unified
picture. Assuming both the protons\footnote{The heavier nuclei which
may play a similar but less important role than protons for $\gamma$-ray
emission are not discussed here.} and electrons are accelerated in
SNRs and the SNRs are the dominant sources of Galactic CRs, one can
derive the electron-to-proton ratio at the source based
on the locally observed spectra of protons and electrons. Then the
$\gamma$-ray emission of SNRs will depend primarily on the environmental
gas density, and the observed diversity of $\gamma$-ray
spectra of SNRs can be attributed to variations of this density.

\section{Electron-proton ratio at the source}

The observational flux of protons at $\sim$GeV is about two
orders of magnitude higher than that of electrons, which implies
the electron-proton ratio at the source should be of the order $1\%$
since the energy loss of electrons is negligible in this low
energy range. This result is well known \citep[e.g.,][]
{1973ICRC....1..634C,1977Ap&SS..46..225A,1994ApJ...426..327L,
2002JPhG...28..359E}. In this section we derive the injection
parameters of the Galactic CRs at the sources by reproducing
the observed spectra on the Earth, considering the
detailed propagation model.

After production at sources, charged energetic
particles propagate diffusively in the random magnetic field of
the Galaxy. The overall convection and reacceleration due to
scattering with the random magnetohydrodynamic waves may also
change the distribution function of CR particles. Furthermore, the
interactions between the CRs and the gas, interstellar radiation
field, and magnetic field, will lead to fragmentation,
catastrophic or continuous energy losses of these particles. The
transport of CRs from the sources to the Earth is generally
complex \citep[][see \cite{2007ARNPS..57..285S} for a recent review
of the CR propagation]{2008MNRAS.386L..20B}.

In this work we limit our study to CR protons and electrons only.
We adopt the GALPROP code\footnote{http://galprop.stanford.edu/}
\citep{1998ApJ...509..212S,1998ApJ...493..694M} version v50p to calculate
the propagation of the CRs. The diffusion-reacceleration frame, without
convection, is assumed. The main propagation parameters are
$D_0=6.59\times10^{28}$ cm$^2$s$^{-1}$, $\delta=0.30$, $v_A=39.2$
km s$^{-1}$ and $z_h=3.9$ kpc, which are derived through the fit to the
B/C, $^{10}$Be/$^9$Be, Carbon and Oxygen data \citep{2011ApJ...729..106T}.
The spatial distribution of the CR sources in the cylindrical coordinates is
\begin{equation}
f(R,z)\propto\left(\frac{R}{R_{\odot}}\right)^{\alpha}\exp\left[-\frac
{\beta(R-R_{\odot})}{R_{\odot}}\right]\exp\left(-\frac{|z|}{z_s}\right),
\end{equation}
where axis symmetry has been assumed and $R_{\odot}=8.5$ kpc is the
distance of solar system from the Galactic center, $z_s\approx 0.2$ kpc is
the scale height of the source distribution, $a=1.25$ and $b=3.56$. Such
a source function is similar to the SNR spatial distribution but tuned
based on the Fermi observations of diffusive Galactic $\gamma$-rays
\citep{2011ApJ...729..106T}.

The injection spectral shape as a function of momentum $p$ (or rigidity)
is assumed to be a broken power-law function
\begin{equation}
q(p)\propto\left\{\begin{array}{cc}
p^{-\alpha_1},& p<p_{\rm br},\\
p^{-\alpha_2},& p\geq p_{\rm br},
\end{array}\right.
\end{equation}
which is identical for both electrons and protons\footnote{For the sake 
of simplicity, we ignore the spectral evolution and potential spectral 
difference between high energy electrons and protons escaping from the 
SNRs \citep{2002JPhG...28..359E}.}. The ratio of the
normalization between electrons and protons, usually called $K_{ep}$, is
taken as a free parameter. The broken power-law injection spectrum is
required to fit the observed CR data \citep{2004ApJ...613..962S,
2011ApJ...729..106T,2012PhRvD..85d3507L} as well as the $\gamma$-ray
data \citep{2010Sci...327.1103A,2009ApJ...706L...1A,
2010ApJ...712..459A,2010ApJ...718..348A,2010ApJ...722.1303A,
2012ApJ...744...80A,2012PhRvL.108e1105N}. It was proposed that strong 
ion-neutral collisions near the shock front may lead to the spectral 
break of accelerated particles around $\sim10$ GeV 
\citep{2005ApJ...624L..37M,2011NatCo...2E.194M}. Alternatively the 
escape effect of particles from/into finite-size region
may also give a break at several GeV \citep{2011MNRAS.410.1577O,
2010MNRAS.409L..35L,2012MNRAS.421..935L}.

\begin{figure}[!htb]
\centering
\includegraphics[width=0.9\columnwidth]{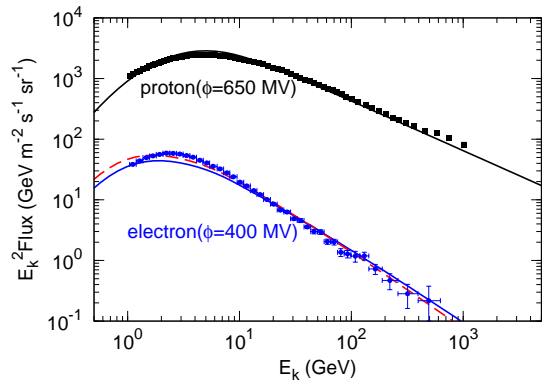}
\caption{The expected fluxes of CR protons and electrons at the Earth,
for the same spectral shape of the injected particles, compared with the
PAMELA observational data \citep{2011PhRvL.106t1101A,2011Sci...332...69A}.
We adopt two parameter settings to calculate the electron spectrum: for
solid line the magnetic field is the canonical one adopted in GALPROP
and $K_{ep}\approx1.3\%$; for dashed line the magnetic field is two times
larger and $K_{ep}\approx1.9\%$.
}
\label{fig:snrpe}
\end{figure}

Compared with the PAMELA observations of the CR proton
\citep{2011Sci...332...69A} and electron spectra \citep{2011PhRvL.106t1101A},
we find $\alpha_1=1.80$, $\alpha_2=2.52$ and $p_{\rm br}c =6$ GeV can
give an acceptable fit to the data, where $c$ is the speed of light\footnote{We are not dedicated to
discuss the spectral hardening of CR nuclei above $\sim200$ GeV reported
by ATIC/CREAM/PAMELA \citep{2007BRASP..71..494P,2010ApJ...714L..89A,
2011Sci...332...69A}, which may imply the superposition of different
source spectra \citep{2011PhRvD..84d3002Y} or a nearby new component of
CRs \citep{2012APh....35..449E,2012MNRAS.421.1209T}.}, as shown in Fig.
\ref{fig:snrpe}.
For energies below $\sim30$ GeV, the solar modulation
with force-field approximation is employed \citep{1968ApJ...154.1011G}.
To fit both the proton and electron spectra with the same injection
spectrum, we need different modulation potentials, as labeled in Fig.
\ref{fig:snrpe}. This may be due to the rest-mass or sign of charge
dependence of the modulation effect \citep{1996ApJ...464..507C}.
To reproduce the absolute fluxes, we find the electron-to-proton ratio
$K_{ep}$ is about $1.3\%$. Such a result could be reproduced with the
numerical simulation \citep{1994ApJ...426..327L}. See
\cite{2002JPhG...28..359E} for more possible explanations.

A better fit to the electron data can be obtained through increasing the
magnetic field\footnote{Note the change of magnetic field may affect
the synchrotron radiation \citep{2011A&A...534A..54S}. What we employ here
is an example to include the uncertainties of the propagation model.
The full discussion of a self-consistent propagation model is beyond the
scope the present work.} by a factor of $2$ (correspondingly
$K_{ep}\approx1.9\%$), as shown by the red-dashed line. Even with an 
identical spectral shape at injection, the electron spectrum is much 
softer than that of protons due to their energy loss in the transport 
processes from the source regions to Earth, which is quite different 
from the scenario explored by \citet{2002JPhG...28..359E}, where SNRs 
produce an electron distribution softer than the proton distribution.

As a comparison, the independent fit to the proton and electron data 
gives $\alpha_1=1.91$, $\alpha_2=2.40$, $p_{\rm br}c=10$ GeV for protons 
\citep{2011ApJ...729..106T}, and $\alpha_1=1.50$, $\alpha_2=2.56$, 
$p_{\rm br}c=3.60$ GeV for electrons \citep{2012PhRvD..85d3507L}. 
Given uncertainties of the propagation parameters, astrophysical inputs, 
solar modulation etc, we consider these spectral fits to be consistent 
with each other.

\section{Gamma-ray emission of supernova remnants}

We now investigate the $\gamma$-ray emission of SNRs adopting
$K_{ep}\sim 1\%$ and the spectral parameters at the source as
derived in the previous section. In general there are three major
components of the $\gamma$-ray emission: IC and bremsstrahlung
radiation by electrons and the $\pi^0$-decay emission by protons.
Therefore we further need the knowledge about the radiation
background and the environmental gas density. For the sake of
simplicity, we assume IC scattering with the cosmic microwave
background radiation only in this work. The scattering with
infrared and optical light may make the IC $\gamma$-ray spectrum
broader \citep{2006ApJ...648L..29P}, but is not expected to affect
the qualitative discussion here, at least for those far away from
the Galactic center. Then the only parameter determining the SNR
$\gamma$-ray spectra will be the gas density.

\begin{figure*}[!htb]
\centering
\includegraphics[width=0.9\columnwidth]{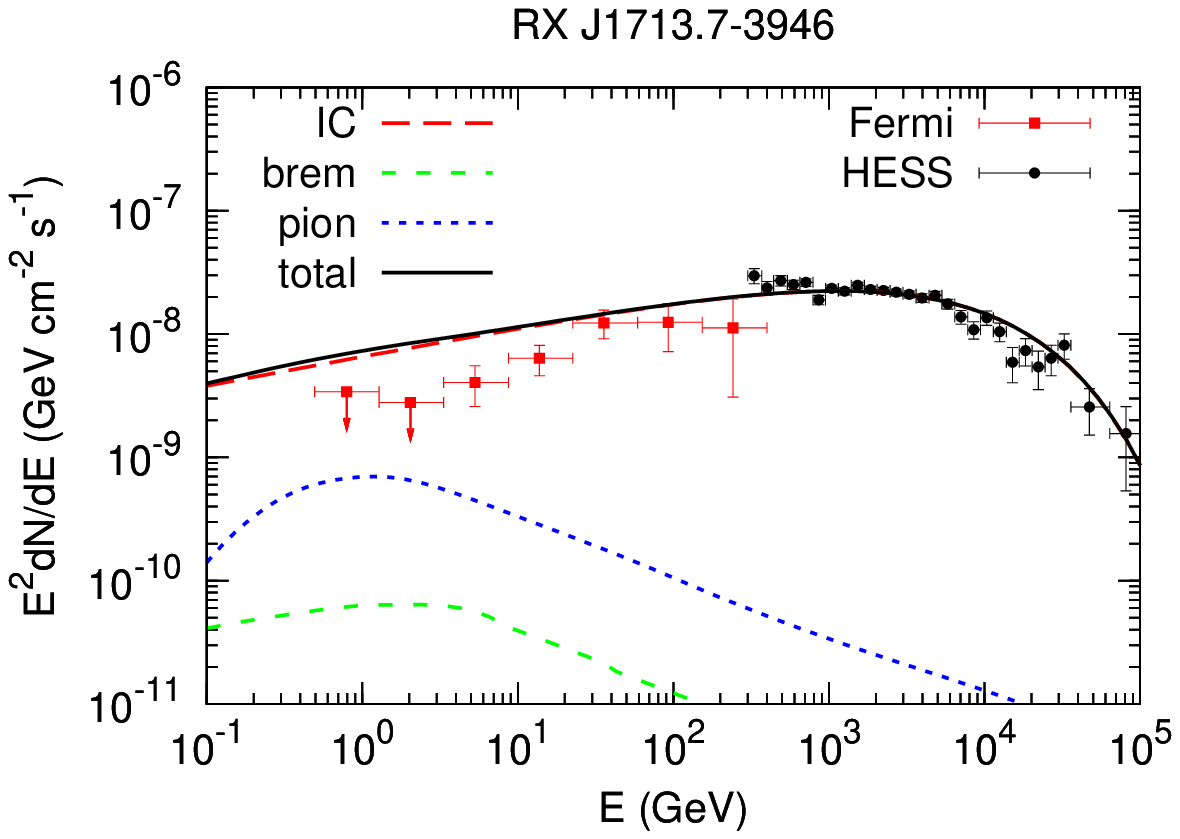}
\includegraphics[width=0.9\columnwidth]{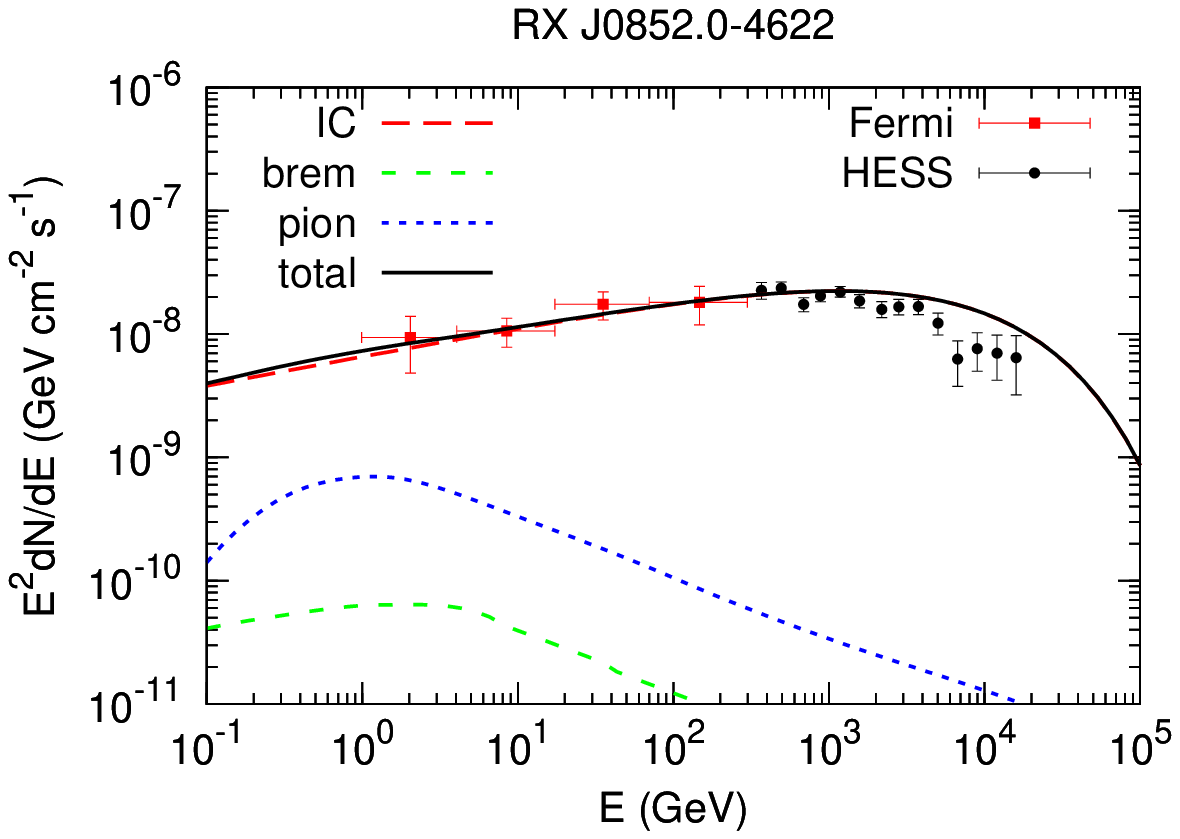}
\caption{Expected $\gamma$-ray spectra for SNRs RX J1713.7-3946 (left) and
RX J0852.0-4622 (right). The gas density is adopted to be $n=0.01$ cm$^{-3}$.
References of the observational data ---
RX J1713.7-3946: Fermi \citep{2011ApJ...734...28A},
HESS \citep{2007A&A...464..235A};
RX J0852.0-4622: Fermi \citep{2011ApJ...740L..51T},
HESS \citep{2007ApJ...661..236A}.
}
\label{fig:low}
\end{figure*}

\begin{figure*}[!htb]
\centering
\includegraphics[width=0.9\columnwidth]{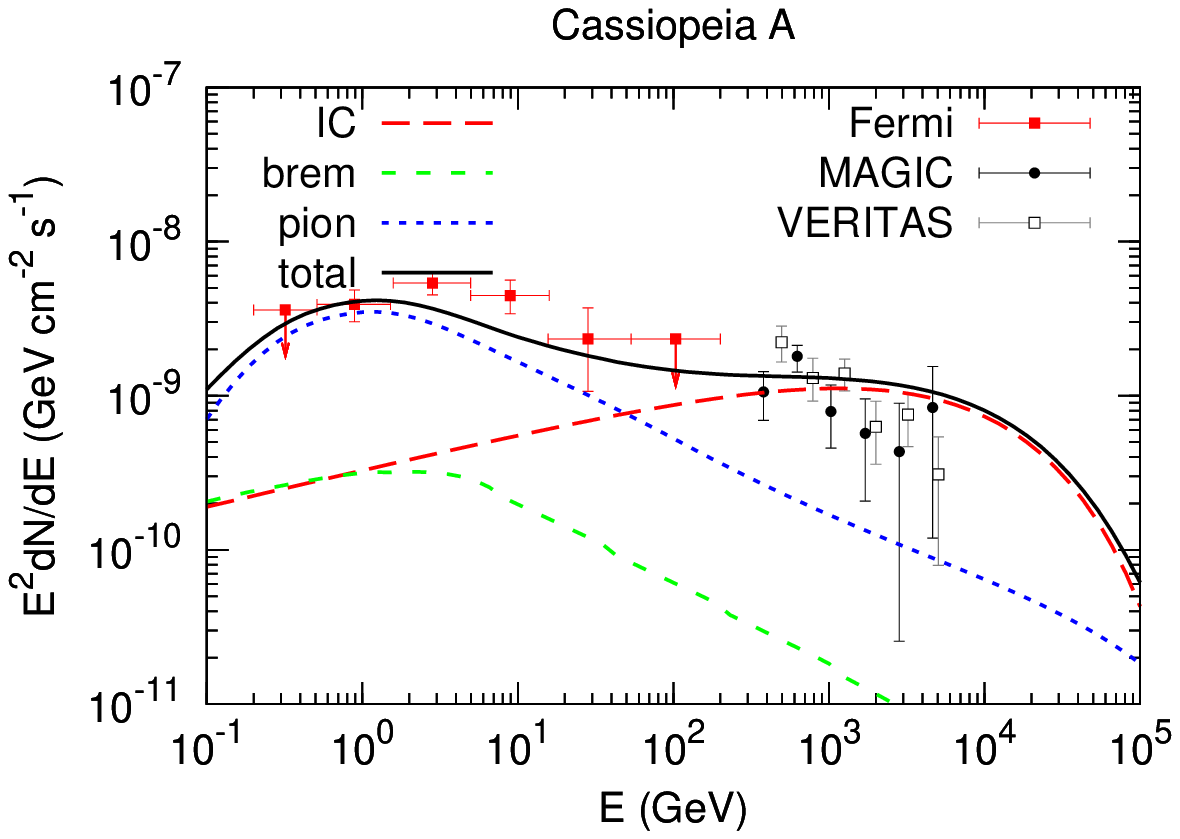}
\includegraphics[width=0.9\columnwidth]{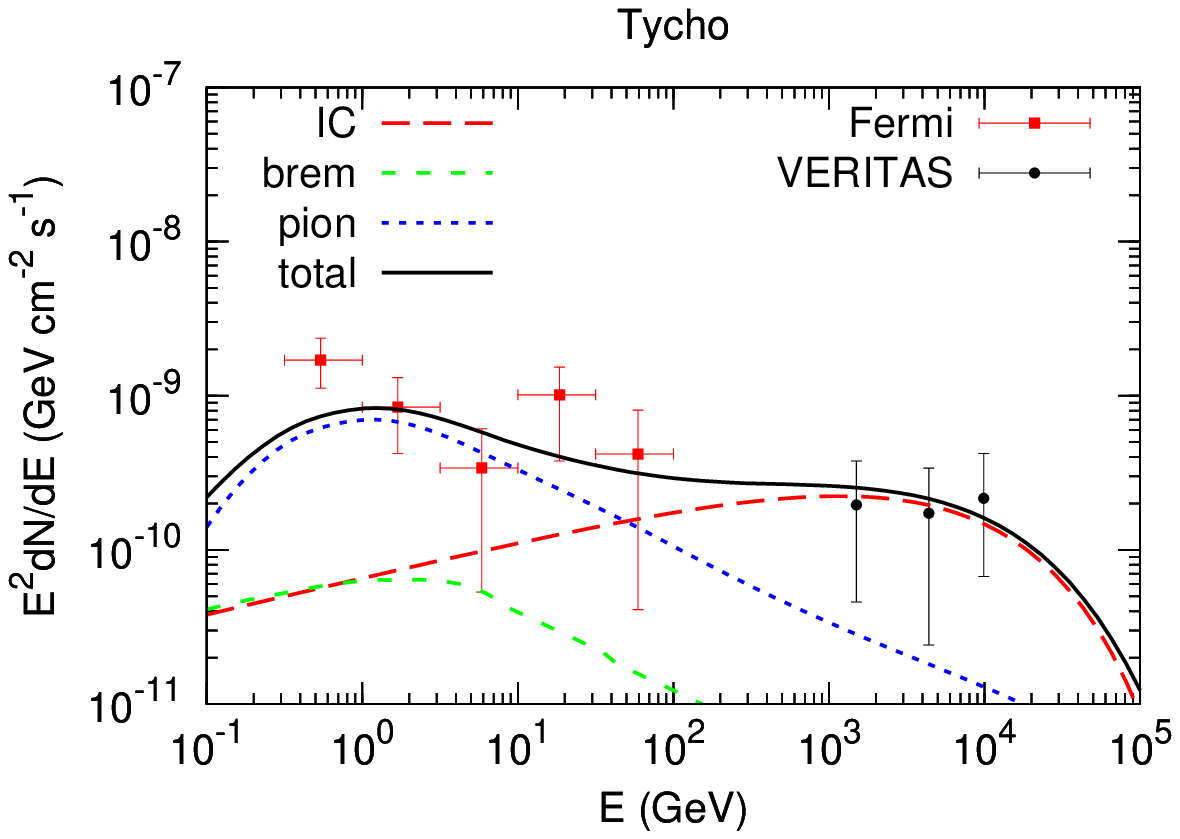}
\caption{Same as Fig. \ref{fig:low} but for Cassiopeia A (left) and
Tycho (right). The gas density is adopted to be $n=1$ cm$^{-3}$.
References of the observational data ---
Cassiopeia A: Fermi \citep{2010ApJ...710L..92A},
MAGIC \citep{2007A&A...474..937A}, VERITAS \citep{2010ApJ...714..163A};
Tycho: Fermi \citep{2012ApJ...744L...2G}, VERITAS \citep{2011ApJ...730L..20A}.
}
\label{fig:medium}
\end{figure*}

\begin{figure*}[!htb]
\centering
\includegraphics[width=0.9\columnwidth]{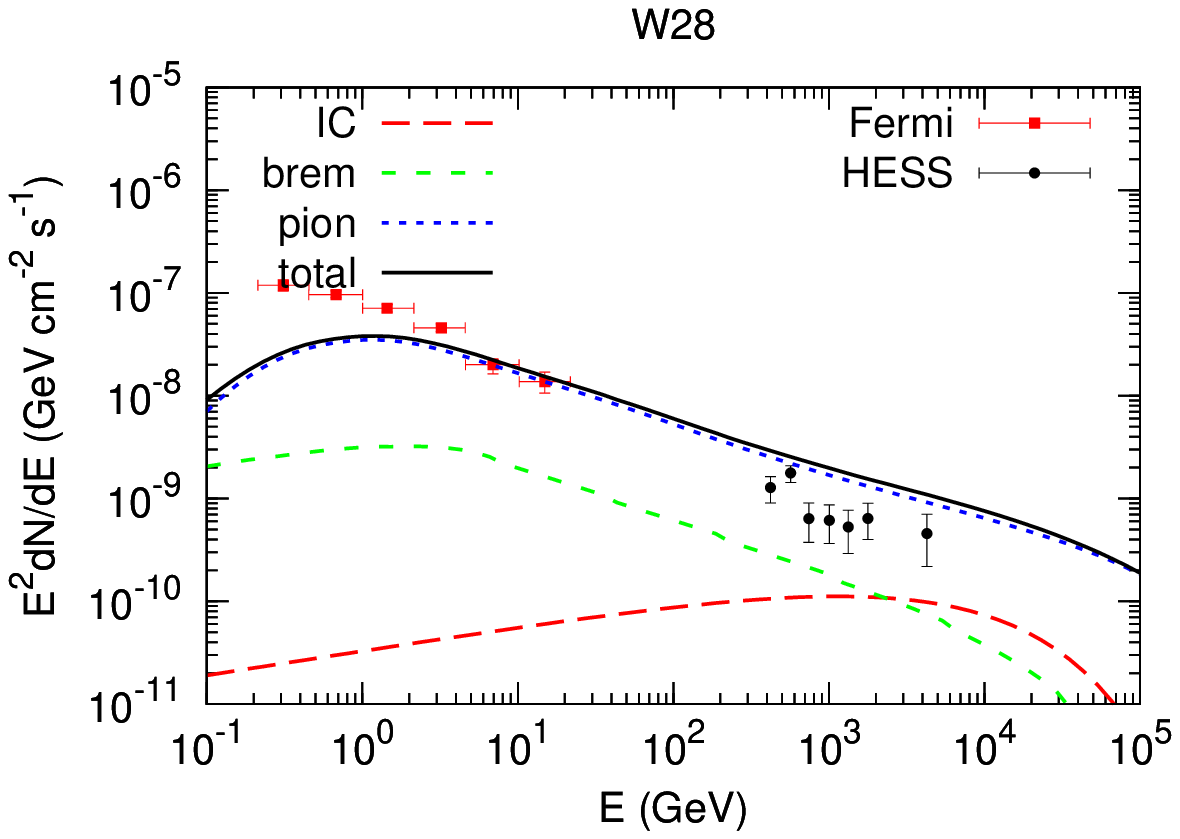}
\includegraphics[width=0.9\columnwidth]{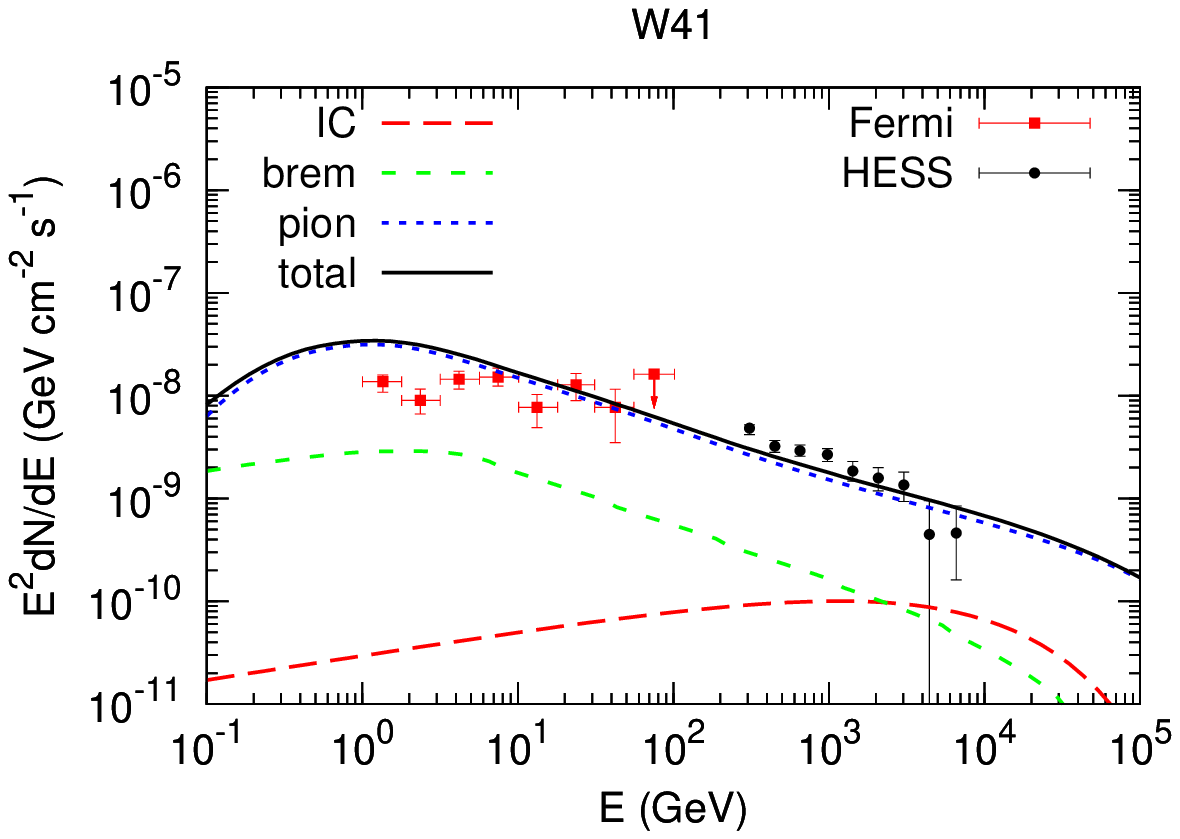}
\includegraphics[width=0.9\columnwidth]{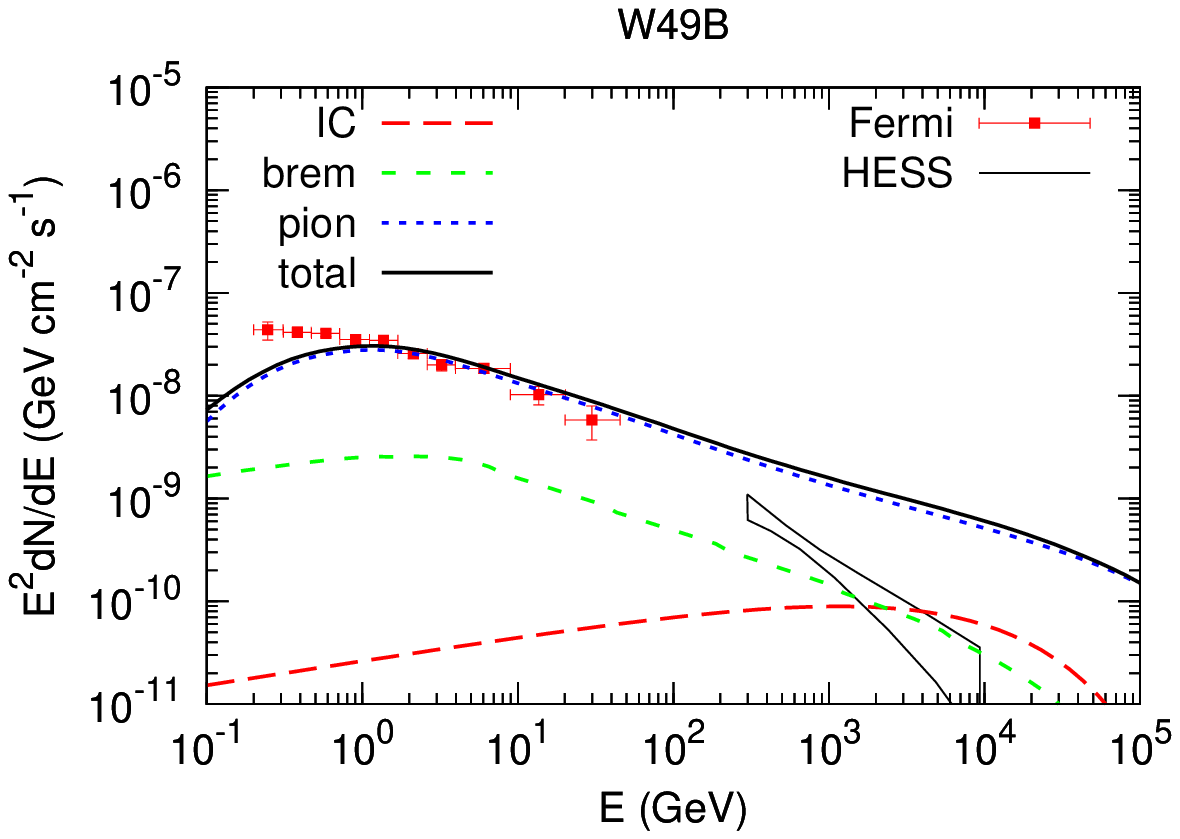}
\includegraphics[width=0.9\columnwidth]{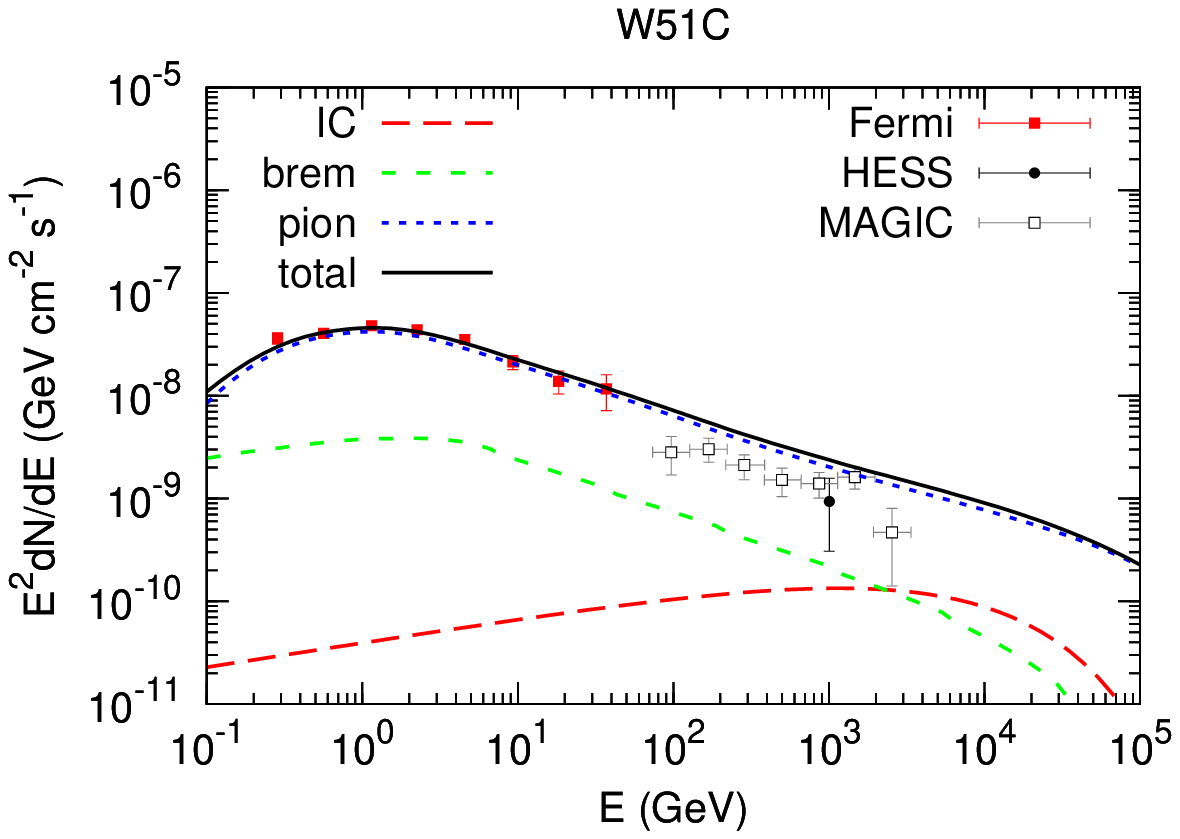}
\includegraphics[width=0.9\columnwidth]{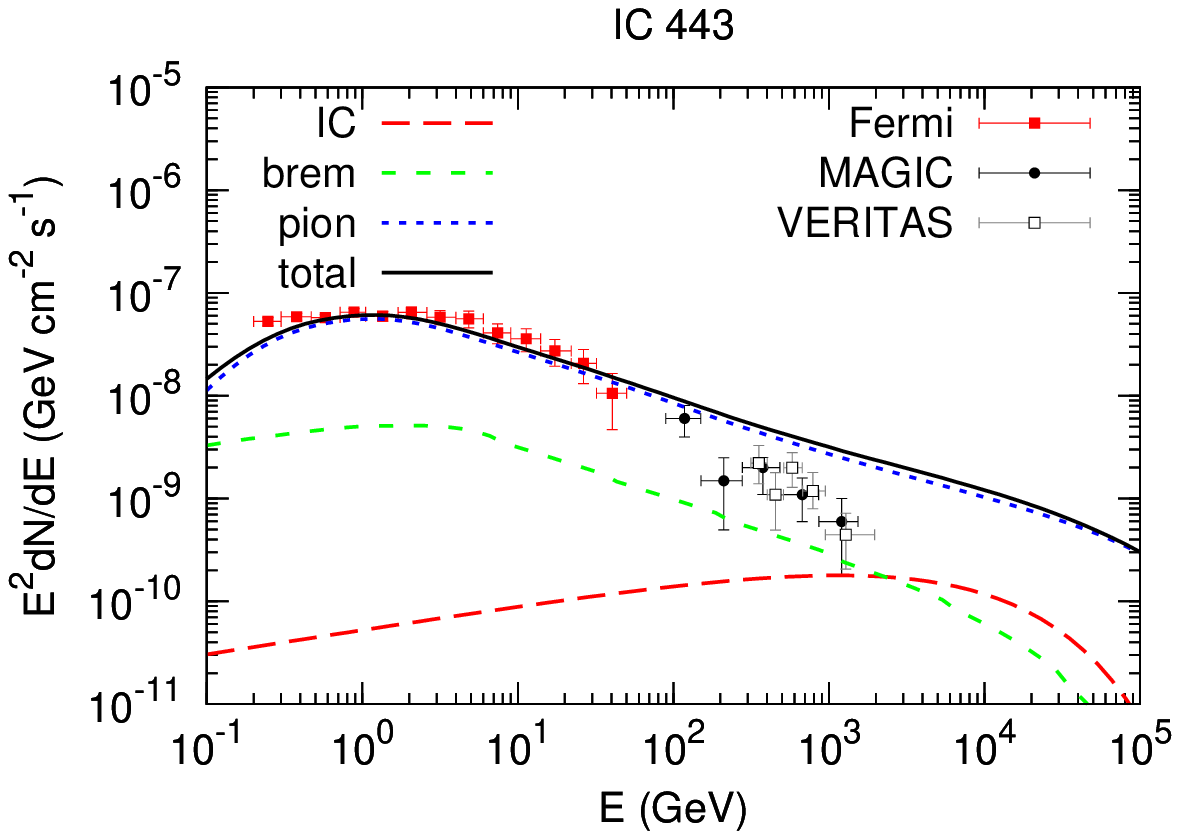}
\includegraphics[width=0.9\columnwidth]{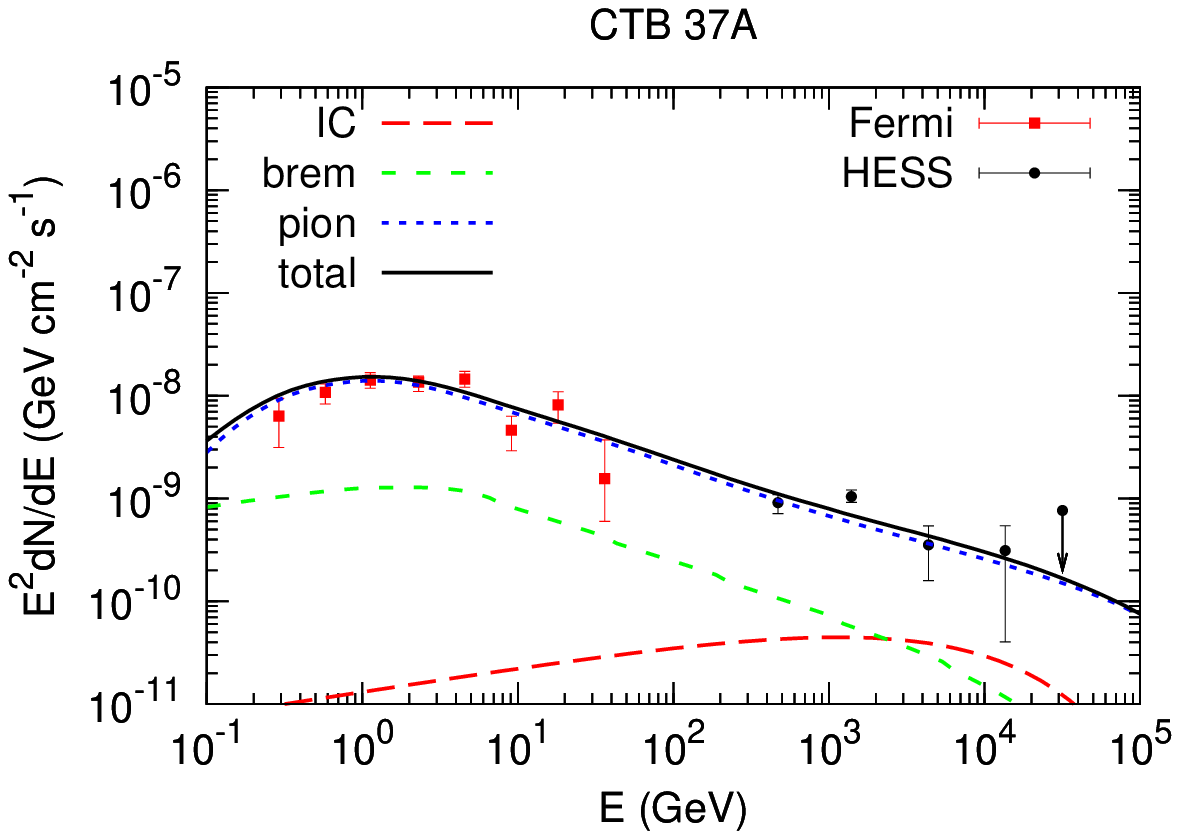}
\includegraphics[width=0.9\columnwidth]{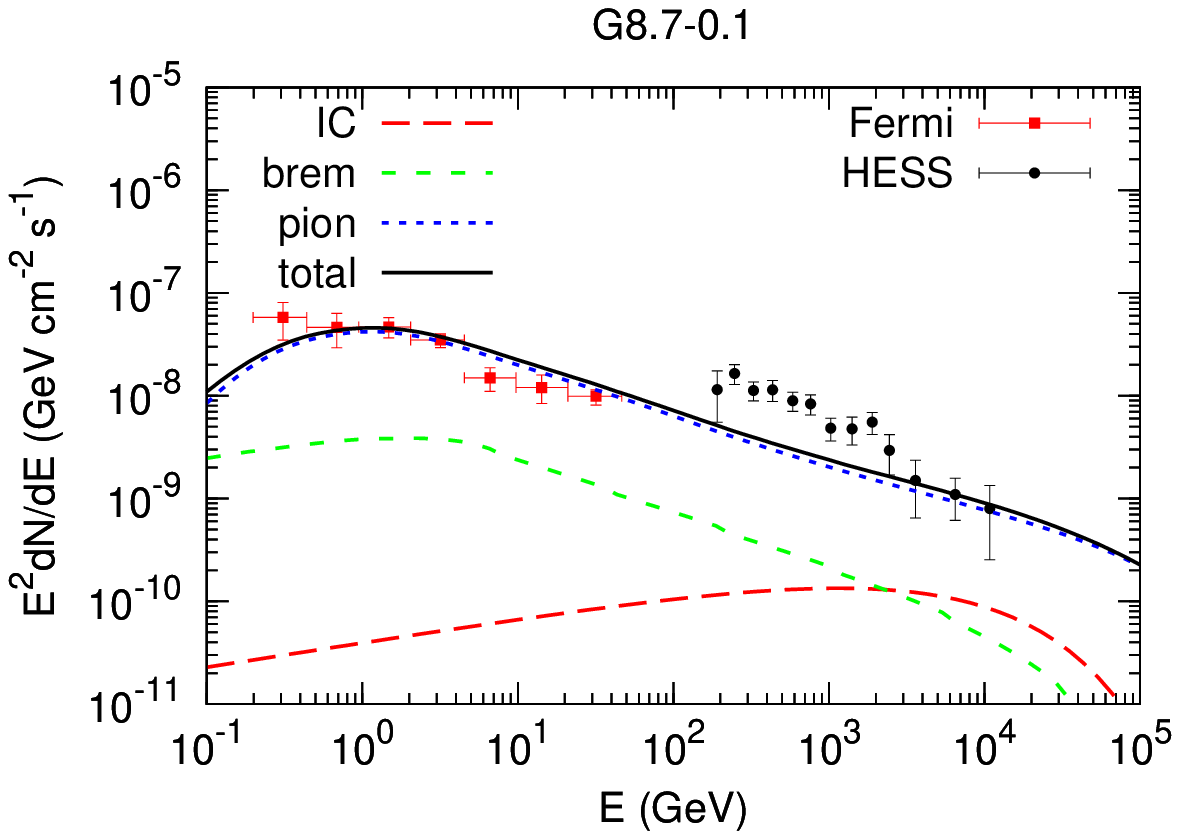}
\includegraphics[width=0.9\columnwidth]{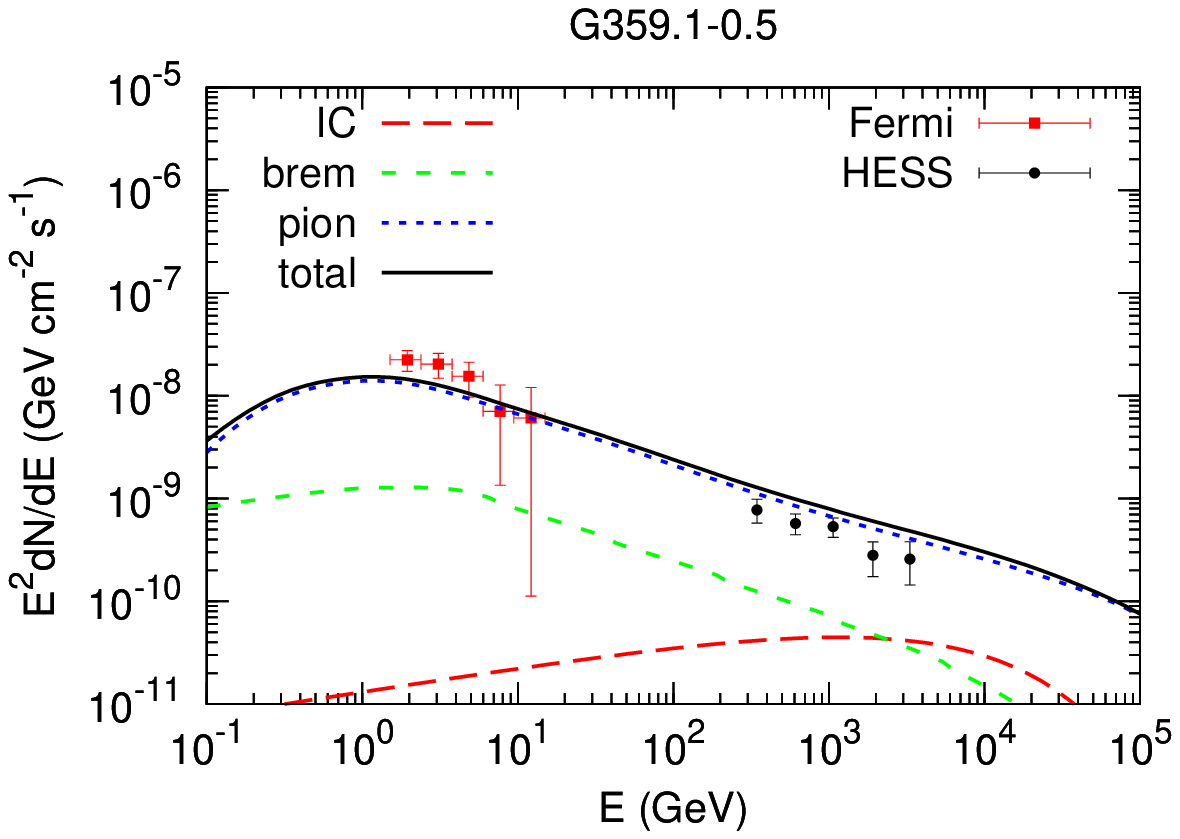}
\caption{Same as Fig. \ref{fig:low} but for SNR-MC interacting systems.
The gas density is adopted to be $n=100$ cm$^{-3}$.
References of the observational data ---
W28: Fermi \citep{2010ApJ...718..348A}, HESS \citep{2008A&A...481..401A};
W41: Fermi \citep{HESS_Fermi_W41}, HESS \citep{HESS_Fermi_W41};
W49B: Fermi \citep{2010ApJ...722.1303A}, HESS \citep{2011arXiv1104.5003B};
W51C: Fermi \citep{2009ApJ...706L...1A}, HESS \citep{HESS_W51C_ICRC},
MAGIC \citep{2011arXiv1110.0950C};
IC 443: Fermi \citep{2010ApJ...712..459A}, MAGIC \citep{2007ApJ...664L..87A},
VERITAS \citep{2009ApJ...698L.133A};
CTB 37A: Fermi \citep{2010ApJ...717..372C}, HESS \citep{2008A&A...490..685A};
G8.7-0.1: Fermi \citep{2012ApJ...744...80A}, HESS \citep{2006ApJ...636..777A};
G359.1-0.5: Fermi \citep{2011ApJ...735..115H}, HESS \citep{2008A&A...483..509A}.
}
\label{fig:high}
\end{figure*}

We classify the SNRs into three groups with low, medium, and high ambient
gas densities. The lack of thermal X-ray emission of RX J1713.7-3946 gives
an upper limit of the gas density of $0.02-0.03$ cm$^{-3}$
\citep{2004A&A...427..199C,2011ApJ...735..120Y}.
For RX J0852.0-4622 ASCA X-ray data implies a gas density $n<0.03(d/1\,
{\rm kpc})^{-1/2}f^{-1/2}$ cm$^{-3}$, with $d$ the distance and $f$ the
filling factor of X-ray emitting volume \citep{2001ApJ...548..814S}.
Although the derivation of the upper limit relies on assumptions such
as the ionization equilibrium and gas temperature, we take them as the
typical examples of low density SNRs. Adopting $n=0.01$ cm$^{-3}$,
together with the CR spectra given by Equation (2) and $K_{ep}\sim1\%$,
we show the calculated $\gamma$-ray spectra of these two SNRs in Fig.
\ref{fig:low}. The overall normalization of the $\gamma$-ray
luminosity is determined by the observational flux of each source.
To be consistent with the high-energy spectral cutoff behavior of both
SNRs, we employ an exponential cutoff term with
$E_c\approx 60$ TeV of the electron spectra. The cutoff might be due
to the balance of acceleration and the cooling in the vicinity of the
SNR. For protons the cutoff could be higher and is assumed not to enter
the energy range discussed here. A remarkable signature of the
$\gamma$-ray spectrum is that, the spectrum is very hard and the
luminosity of TeV emission is higher than that of the GeV emission.
The $\gamma$-ray emission of these SNRs is IC dominated.

Some relatively young SNRs have a moderate gas density, such as Cassiopeia
A and Tycho. The average gas density of Cassiopeia A is estimated to be about
$4.4$ cm$^{-3}$ \citep{2010ApJ...720...20A}. For Tycho an upper limit
$n<0.6$ cm$^{-3}$ was derived from the absence of thermal X-ray emission
from the bright outer rim of the remnant \citep{2007ApJ...665..315C}. 
The density in the inner region of the remnant can be much higher. Here
we take these two SNRs as examples of medium density sample and adopt 
$n=1$ cm$^{-3}$ \footnote{Although these sources clearly show complex 
source structure with multiple emission zones\citep{2012ApJ...749L..26A}, 
in this paper we still consider the simple one zone emission model to 
capture the dominant features.}. The calculated spectra are shown in 
Fig. \ref{fig:medium}. We can see that for this kind of sources the 
GeV emission is $\pi^0$-decay dominated and the TeV emission is IC 
dominated. The luminosities in GeV and TeV bands are comparable in this case.

Finally we discuss the case with a high density, typically for the SNR-MC
interacting systems. The gas density in the MCs can easily reach
$10^2-10^3$ cm$^{-3}$. As an illustration we adopt $n=100$ cm$^{-3}$
in this study. The expected $\gamma$-ray spectra together with the
GeV-TeV observational data of eight SNR-MC systems are shown in
Fig. \ref{fig:high}. In this case the GeV-TeV $\gamma$-ray emission
is $\pi^0$-decay dominated, and the $\gamma$-ray spectrum is
generally very soft.

We may need to estimate the impact of energy losses in the dense
clouds on the CR proton spectrum. For $n=100$ cm$^{-3}$
ISM, the ionization energy loss time scale of protons is about $10^6$
yr at $1$ GeV \citep{1998ApJ...509..212S}, and the pion-production energy
loss time scale is about $10^5$ yr. For most of these SNRs the ages are
estimated to be less than several tens kyr, so we expect that for the
energy range interested in this work (GeV-TeV) the energy losses of the
protons are not important. For the low energy particles (several tens MeV)
and high energy electrons ($\gtrsim$ TeV), the energy losses may be important
and need to be considered in future modeling. As for the contribution
to $\gamma$-rays from the secondary $e^{\pm}$, it is only important
for $E_{\gamma}<10$ MeV compared with the $pp$ induced $\pi^0$-decay
component, even for very old SNR \citep{2008MNRAS.384.1119F}.

\begin{figure}[!htb]
\centering
\includegraphics[width=0.9\columnwidth]{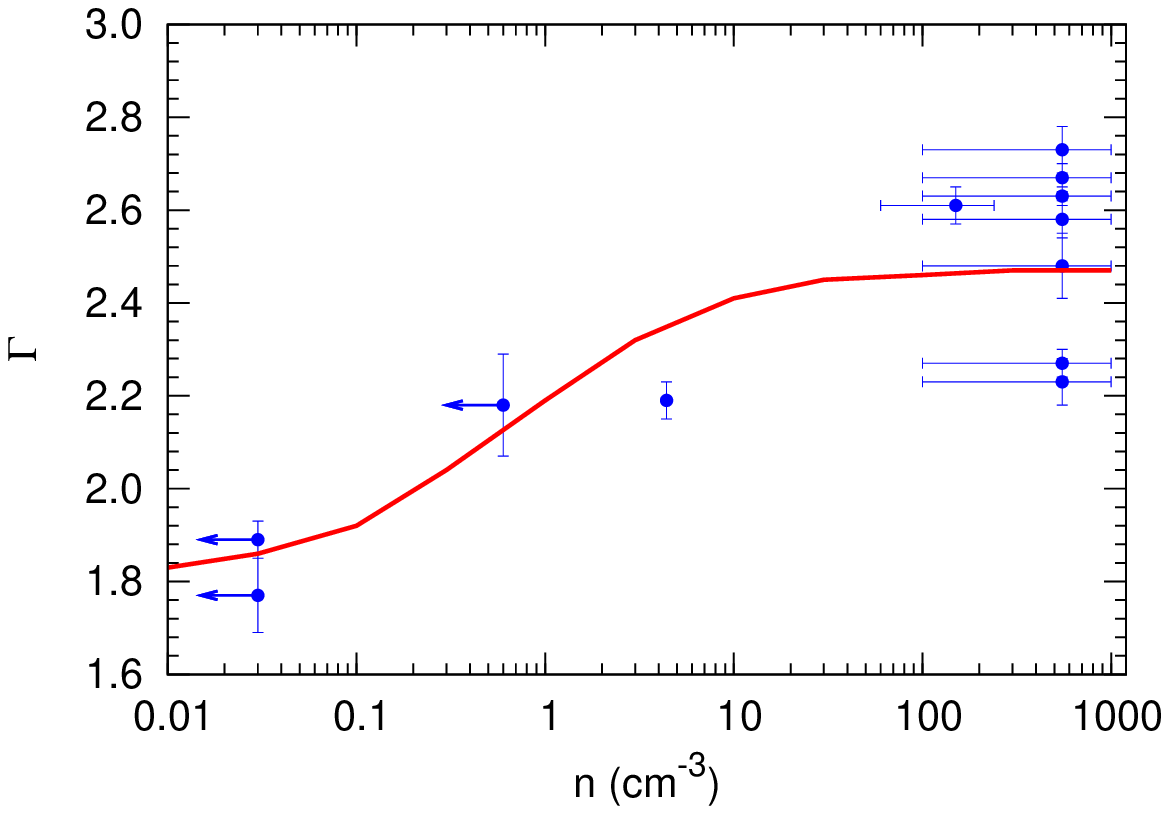}
\caption{The photon index $\Gamma$ (between 1 GeV and 1 TeV) versus the
gas density $n$ of the $12$ SNRs studied in this work. The solid line is
the model expected result.
}
\label{fig:nspec}
\end{figure}

In Fig. \ref{fig:nspec} we show the relation between the photon index
$\Gamma$ and the gas density $n$, for the SNRs studied in this work.
The parameters of the SNRs are compiled in Table \ref{table:para}. The
photon index $\Gamma$ is fitted using the observational data between $1$
GeV and $1$ TeV, with single power-law function. For the SNR-MC interacting
system whose gas density is not well known we assume a value of $10^2-10^3$
cm$^{-3}$. A trend showing the correlation between $\Gamma$ and $n$
can be seen from Fig. \ref{fig:nspec}. We also show the theoretical
expected result based on the unified model with the solid line. The model
is consistent with the observational data.

\begin{table*}[!htb]
\centering
\caption{Parameters of the SNRs: Name, R.A., Dec., distance, age, gas
density, photon index (between 1 GeV and 1 TeV) and references.
}
\begin{tabular}{cccccccc}
\hline
\hline
Name & R.A. & Dec. & $d$ (kpc) & Age (kyr) & $n$ (cm$^{-3}$) & $\Gamma$ & Ref.\\
\hline
RX J1713.7-3946 & $17^h13^m50^s$ & $-39^{\circ}45'$ & $1.0$ & $1.6$ & $<0.03$ & $1.77\pm0.08$ & 1,2 \\
RX J0852.0-4622 & $08^h52^m00^s$ & $-46^{\circ}20'$ & $0.75$ & $1.7-4.3$ & $<0.03$ & $1.89\pm0.04$ & 3,4 \\
Cassiopeia A & $23^h23^m26^s$ & $+58^{\circ}48'$ & $3.4$ & $0.32$ & $4.4$ & $2.19\pm0.04$ & 5-8 \\
Tycho & $00^h25^m18^s$ & $+64^{\circ}09'$ & $2.5-3.0$ & $0.44$ & $<0.6$ & $2.18\pm0.11$ & 9-12 \\
W28 & $18^h00^m30^s$ & $-23^{\circ}26'$ & $2.0$ & $35-150$ &  --- & $2.67\pm0.03$ & 13,14 \\
W41 & $18^h34^m45^s$ & $-08^{\circ}48'$ & $3.9-4.5$ & $60-200$ & --- & $2.27\pm0.03$ & 15,16 \\
W49B & $19^h11^m08^s$ & $+09^{\circ}06'$ & $8-12$ & $1-4$ & --- & $2.73\pm0.05$ & 17,18 \\
W51C & $19^h23^m50^s$ & $+14^{\circ}06'$ & $6.0$ & $30$ & --- & $2.58\pm0.04$ & 19-21 \\
IC 443 & $06^h17^m00^s$ & $+22^{\circ}34'$ & $0.7-2.0$ & $3-30$ & $60-240$ & $2.61\pm0.04$ & 22-24 \\
CTB 37A & $17^h14^m06^s$ & $-38^{\circ}32'$ & $6.3-9.5$ & --- & --- & $2.48\pm0.07$ & 25-27 \\
G8.7-0.1 & $18^h05^m30^s$ & $-21^{\circ}26'$ & $4.8-6.0$ & $15-28$ & --- & $2.23\pm0.05$ & 28,29 \\
G359.1-0.5 & $17^h45^m30^s$ & $-29^{\circ}57'$ & $7.6$ & $>10$ & --- & $2.63\pm0.02$ & 30,31 \\
\hline
\vspace{1mm}
\end{tabular}\\
Reference --- (1) \cite{2011ApJ...734...28A}; (2) \cite{2007A&A...464..235A};
(3) \cite{2011ApJ...740L..51T}; (4) \cite{2007ApJ...661..236A};
(5) \cite{2010ApJ...710L..92A}; (6) \cite{2007A&A...474..937A};
(7) \cite{2010ApJ...714..163A}; (8) \cite{2006ApJ...645..283F};
(9) \cite{2012ApJ...744L...2G}; (10) \cite{2011ApJ...730L..20A};
(11) \cite{2011ApJ...729L..15T}; (12) \cite{2007ApJ...665..315C};
(13) \cite{2010ApJ...718..348A}; (14) \cite{2008A&A...481..401A};
(15) \cite{HESS_Fermi_W41}; (16) \cite{2008AJ....135..167L};
(17) \cite{2010ApJ...722.1303A}; (18) \cite{2011arXiv1104.5003B};
(19) \cite{2009ApJ...706L...1A}; (20) \cite{HESS_W51C_ICRC};
(21) \cite{2011arXiv1110.0950C}; (22) \cite{2010ApJ...712..459A};
(23) \cite{2007ApJ...664L..87A}; (24) \cite{2009ApJ...698L.133A};
(25) \cite{2010ApJ...717..372C}; (26) \cite{2008A&A...490..685A};
(27) \cite{2012MNRAS.421.2593T}; (28) \cite{2012ApJ...744...80A};
(29) \cite{2006ApJ...636..777A}; (30) \cite{2011ApJ...735..115H};
(31) \cite{2008A&A...483..509A}.
\label{table:para}
\end{table*}

It is encouraging that the results in Figs. \ref{fig:low}-\ref{fig:high}
show rough consistence with the observational data, in support of our
relatively simple interpretation of the complicated $\gamma$-ray spectral
behaviors of SNRs. Note here we have not tried to precisely fit the
observational data because variations of the spectral and environmental
parameters are expected for different sources \citep{2010A&A...510A.101F,
2011APh....35...33Y,2011PhRvD..84d3002Y}. With slight adjustment of these
source parameters we can easily get better fit to the data
\citep{2012MNRAS.421..935L}.

\section{Conclusion and discussion}

In this work we propose a unified model to explain the $\gamma$-ray
emission of SNRs. In the model, by assuming that SNRs produce identical high energy electron and proton spectral shape and are the major
sources of the low energy CRs (below the ``knee''), the electron-to-proton
number ratio $K_{ep}$ at the sources is derived to be $\sim1\%$ according
to a realistic CR propagation model described with the GALPROP code.
With such a $K_{ep}$ value we calculate the expected $\gamma$-ray spectra
for various SNRs with different environmental parameters. Qualitatively
the observed diversity of $\gamma$-ray spectra of different SNRs can be
naturally understood with different gas densities. For low density
environments the $\gamma$-ray emission is IC dominated, while for
high density environments the $\gamma$-ray emission is $\pi^0$-decay
dominated. The model predicts that $\gamma$-ray spectra in low density
environments are general harder than those of SNR-MC interaction systems. Since strong thermal emission is expected from shocked dense media, we expect relatively weaker thermal emission from remnants with harder $\gamma$-ray spectra than those with softer $\gamma$-ray spectra.
Such a simple, self-consistent model, if further validated by
observations, supports the SNR-origin of the low-energy CRs.

The age could also be a parameter affecting the $\gamma$-ray emission
of SNRs, and it can be coupled with the density parameter. For example,
considering the progenitors, the density is in the intermediate range for
very young remnants, lowest for middle-age SNRs, and highest for very old
remnants \citep{2005ApJ...630..892D}. An improved unified model of
course needs to consider more factors, including the effect of multiple 
emission zones as observed in some sources \citep{2012ApJ...749L..26A}, 
which will affect the detailed spectral fit.

In general, for $K_{ep}\sim1\%$, the hadronic component will
always dominate the bremsstrahlung component, which is distinguishable
from the model that bremsstrahlung may dominate the $\gamma$-ray emission
\citep{2012ApJ...749L..26A}, and could be tested by the observation of
$\gamma$-rays in lower energy range ($<100$ MeV).

When calculating the $\gamma$-ray emission of the SNRs, we employ the
particle spectra which are the same as the injection spectra giving rise to
the locally observed ones. However, we should keep in mind that the CR
spectra accelerated by the source at specific epoch may be different from
that injected into the Milky Way, which should be the time integrated
spectrum \citep{2010APh....33..160C}. A more detailed modeling may need
to take into account the evolution history of the SNRs \citep[e.g.,][]
{1997ApJ...490..619S,2008ARA&A..46...89R,2008ApJ...686..325L,
2008MNRAS.384.1119F,2012ApJ...751...65F,2012ApJ...745..140Y}.
Furthermore, the electron-to-proton ratio is assumed to be a constant in
this work, independent of the sources. Such an assumption can break the
degeneracy between medium density $n$ and $K_{ep}$. It is a strong
assumption but is not in conflict with observations. Detailed study of
individual sources is needed to verify this assumption.

The magnetic field of the shocked emission region is an important 
parameter but does not appear explicitly in our model. However, for the 
multi-wavelength modeling including the synchrotron emission component, 
the average magnetic field might be well-determined via a detailed 
spectral fit \citep{2010A&A...517L...4F}. Indeed with the increase of 
the magnetic field, the IC component will be suppressed for a given 
synchrotron flux due to the decrease of the number of energetic electrons. 
The gas density (or number of CR protons and therefore $1/K_{ep}$) needs 
to be increased in this case to account for the observed $\gamma$-ray 
flux. The $\gamma$-ray emission is likely dominated by pionic emission 
due to higher gas density or proton flux in this case, and it will be 
difficult to detect the IC component. Another effect for the high magnetic 
field is the cooling of accelerated electrons via synchrotron emission. 
Thus the electron spectrum should be in contradiction with our assumption 
of a unified spectral shape. Except for this particular case our study 
will remain valid as far as this strong magnetic field does not dominate 
the overall particle acceleration in SNRs. 
In cases where IC dominates the $\gamma$-ray emission, the magnetic 
field can be directly derived through the radio to X-ray emission of 
the SNRs. The magnetic field is usually weak ($\sim 10\mu$G) with an
energy density only a factor of a few times higher than that of the 
background photons, according to the synchrotron X-ray to IC $\gamma$-ray 
luminosity ratio \citep{2008ApJ...683L.163L}. \citet{1983A&A...125..249L} 
showed that with such a weak field, SNRs can barely accelerate charged 
particles up to 10$^{15}$ eV. Although the discussion in this work may 
still hold ($E\lesssim100$ TeV) even in the case of weak magnetic field, 
for CRs to be accelerated up to the knee via the diffusive shock 
acceleration, magnetic fields need to be amplified further by some 
mechanisms \citep[e.g.,][]{2004MNRAS.353..550B,2012ApJ...747...98G}.
Inhomogeneity of the magnetic field and multi-zone acceleration scenario 
\citep{2012ApJ...749L..26A} may alleviate challenges to the weak field case.
Better determination of the magnetic field, together with the gas density
parameter, will provide crucial tests of the assumptions made in this work.

Finally, the radio observations of SNRs indicate that the electron
spectrum is about $E^{-2}$ with a remarkable dispersion
\citep{1976MNRAS.174..267C,1985ICRC....9..543B,2002cra..book.....S},
which is not exactly the same as that inferred in Sec. 2. We also 
expect dispersion of other parameters characterizing the energetic 
particle distribution, which may be used to improve the fit to the 
observed $\gamma$-ray spectra. The present work gives a zero-order 
approximation to the problem of the SNR $\gamma$-ray emission and 
the origin of CRs. Some common features of the $\gamma$-ray emission 
of different population of SNRs are revealed. Further works about 
the details may be helpful to refine the present model.







\acknowledgments
We thank A. W. Wolfendale for useful comments on the manuscript and the 
anonymous referee for helpful suggestions.
This work is supported by the Natural Sciences Foundation of China under
grant Nos. 11075169, 11105155, 11143007, and 11173064, the 973 project
under grant No. 2010CB833000 and Chinese Academy of Sciences under grant
No. KJCX2-EW-W01.

\bibliographystyle{apj}
\bibliography{/home/yuanq/work/cygnus/tex/refs}

\end{document}